\RequirePackage{lineno}
\documentclass[prl,preprintnumbers,twocolumn,amsmath,superscriptaddress,amssymb]{revtex4}

\usepackage{graphicx}
\usepackage{dcolumn}
\usepackage{bm}
\usepackage{verbatim}
\usepackage{xcolor}

\begin{document}


\title{Nuclear Fusion in Laser-Driven Counter-Streaming Collisionless Plasmas}

\def\sjtu {Department of Physics and Astronomy, Shanghai Jiao Tong University, Shanghai, 200240, China}
\def\iop {Laboratory of Optical Physics, Institute of Physics, Chinese Academy of Sciences, Beijing 100190, China}
\def\imp {Institute of Modern Physics, Chinese Academy of Sciences, Lanzhou, 730000, China}
\def\ciae {Department of Nuclear Physics, China Institute of Atomic Energy, Beijing, 102413, China}
\def\korea {Center for Relativisitic Laser Science, IBS, Gwangju 61005, Korea}
\def\nao {National Astronomical Observatories, Chinese Academy of Sciences, Beijing 100012, China}
\def\ifsa {IFSA Collaborative Innovation Center, Shanghai Jiao Tong University, Shanghai 200240, China}
\def\siom{Shanghai Institute of Optics and Fine Mechanics, Chinese Academy of Sciences, Shanghai 201800, China}

\author{Xiaopeng Zhang} \affiliation{\sjtu}
\author{Jiarui Zhao}  \affiliation{\iop}
\author{Dawei Yuan} \affiliation{\nao}
\author{Changbo Fu}
\email[Corresponding author:] {cbfu@sjtu.edu.cn} \affiliation{\sjtu}
\author{Jie Bao}  \affiliation{\ciae}
\author{Liming Chen} \affiliation{\iop}\affiliation{\ifsa}
\author{Jianjun He} \affiliation{\nao}\affiliation{\imp}
\author{Long Hou} \affiliation{\ciae}
\author{Liang Li} \affiliation{\sjtu}
\author{Yanfei Li} \affiliation{\iop}
\author{Yutong Li} 
\email[Corresponding author:] {ytli@iphy.ac.cn} \affiliation{\iop}\affiliation{\ifsa}
\author{Guoqian Liao} \affiliation{\iop}
\author{Yongjoo Rhee} \affiliation{\korea}
\author{Yang Sun} \affiliation{\sjtu}\affiliation{\ifsa}
\author{Shiwei Xu} \affiliation{\imp}
\author{Gang Zhao}  \affiliation{\nao}
\author{Baojun Zhu}  \affiliation{\iop}
\author{Jianqiang Zhu}  \affiliation{\siom}
\author{Zhe Zhang}  \affiliation{\iop}
\author{Jie Zhang} 
\affiliation{\sjtu}\affiliation{\ifsa}


\date{\today}

\begin{abstract}
Nuclear fusion reactions are the most important processes in nature to power stars and produce new elements, 
and lie at the center of the understanding of nucleosynthesis in the universe.
It is critically important to study the reactions in full plasma environments that are close to true astrophysical conditions.
By using laser-driven counter-streaming collisionless plasmas,
we studied the fusion D$+$D$\rightarrow n +^3$He in a Gamow-like window around 27 keV.
The results show that astrophysical nuclear reaction yield can be modulated significantly by the self-generated
electromagnetic fields and the collective motion of the plasma. 
This plasma-version mini-collider may provide a novel tool for
 studies of astrophysics-interested nuclear reactions 
 in plasma with tunable energies in earth-based laboratories.
\end{abstract}
\maketitle

\maketitle

Collisionless plasma (CLP) exists in many astrophysical
environments. Well-known examples of CLPs are widely-found
collisionless shockwaves in supernovae remnants, in gamma-ray
bursts, and in solar winds, etc\cite{Biskamp1973}.
In these special environments,
nucleosynthesis cross sections could be significantly different from that in usual
cases due to the following facts. 
Firstly, the energy distribution of the particles in a CLP may be far from thermal equilibrium. 
Due to the ``collisionless'' features\cite{Bret2015}, 
some charged particles in it can be continuously accelerated in a large scale
without  losing their energy too much through scattering \cite{blandford1987PhysRep}, 
and the system can keep in non-thermal equilibrium for long time.
Secondly, the self-generated macro-scale electromagnetic field,
originating from the effects such as the Biermann battery effect and
the Weibel instability \cite{ryu2012magnetic}, can affect the motion
of the particles, and thus their nuclear reaction yield.  
Lastly, the reaction yield could be significantly modified by the so-called
electron screening effect \cite{rolfs1988cauldrons}. For nuclei in
the normal atomically-bound states, 
their decay properties and reaction rates can be completely different from
those in plasma environments \cite{Beta-decay2011,DD-e-Screening2001}. 
However, almost all nuclear parameters used as
astrophysical inputs are traditionally measured under non-plasma
environments. Obviously, creation of plasma conditions in
terrestrial laboratories for studying astrophysical nuclear
reactions is critically important, which may help solving some
long-standing nucleosynthesis puzzles, as for instance, the puzzles
on $^{26}$Al \cite{INSPEC5247128Al26} and $^{6,7}$Li abundance
\cite{cyburt2016BBN}.

The development of high-intensity laser technologies makes it
possible to create plasma environments for nuclear studies in earth-based laboratories 
\cite{Remington1999},  for example, the coulomb
explosion method \cite{DDn-Nature1999, Barbui.PRL.DD}, the inertial confined fusion method
\cite{DT2012Laser}, and the double lasers method
\cite{2laser-labaune2013fusion} etc. However, these methods have limits like untunable energy, large facility needed etc.

Here we report the studies of D(d,n)$^3$He
 in a CLP by using laser-driven head-on-head collision of
plasma streams\cite{jets-li2013structure, CountingSteaming-Jap-kuramitsu2009jet,
collisional-Jets-ryutov2011, B-Self-org-kugland2012,
Char-Counter-Jet-ross2012, Jet-shocks-ryutov2012intra,
huntington2015observation}. 
Because of the head-on-head collision, there is an
enhancement of the center-of-mass (c.m.) energy by a factor of four.
The reaction yields are thereby significantly enhanced as the
reaction cross-section increases exponentially with the c.m. energy.
We show that this kind of mini-version plasma collider
\cite{Zhao2016} could be an ideal method
for studying nuclear reactions in plasmas at earth-based laboratories
with features like tunable energies to cover the astrophysical Gamow window
\cite{rolfs1988cauldrons}.

The experiment was carried out at the Shenguang II laser
facility, the National Laboratory on High Power Lasers and Physics
in Shanghai, China. 
The experimental setup is shown in Fig.\ref{fig.setup}. 
There were eight laser beams at this facility, 
each of which could deliver an energy of about 250 J with the pulse width of 1 ns
at wavelength of 351 nm (3$\omega$). 
Two targets were located at the center of the laser chamber. 
Both of the targets had 0.5$\times$0.5mm$^2$ sized copper bases 
which were coated with 5 or 10 $\mu$m thick deuterated hydrocarbon (CD$_{1.29}$) layers and a separation of 4.4 mm between them
\cite{ZhaoJR2015}.
The main lasers were arranged as two sets (4+4) and each set had four lasers focusing on one of the
targets. The diameter of the focus spots was about 150 $\mu$m,
producing a laser intensity of about $6\times 10^{15}$~W/cm$^2$.

\begin{figure*}
 \centering
 \includegraphics[width=0.8\textwidth]{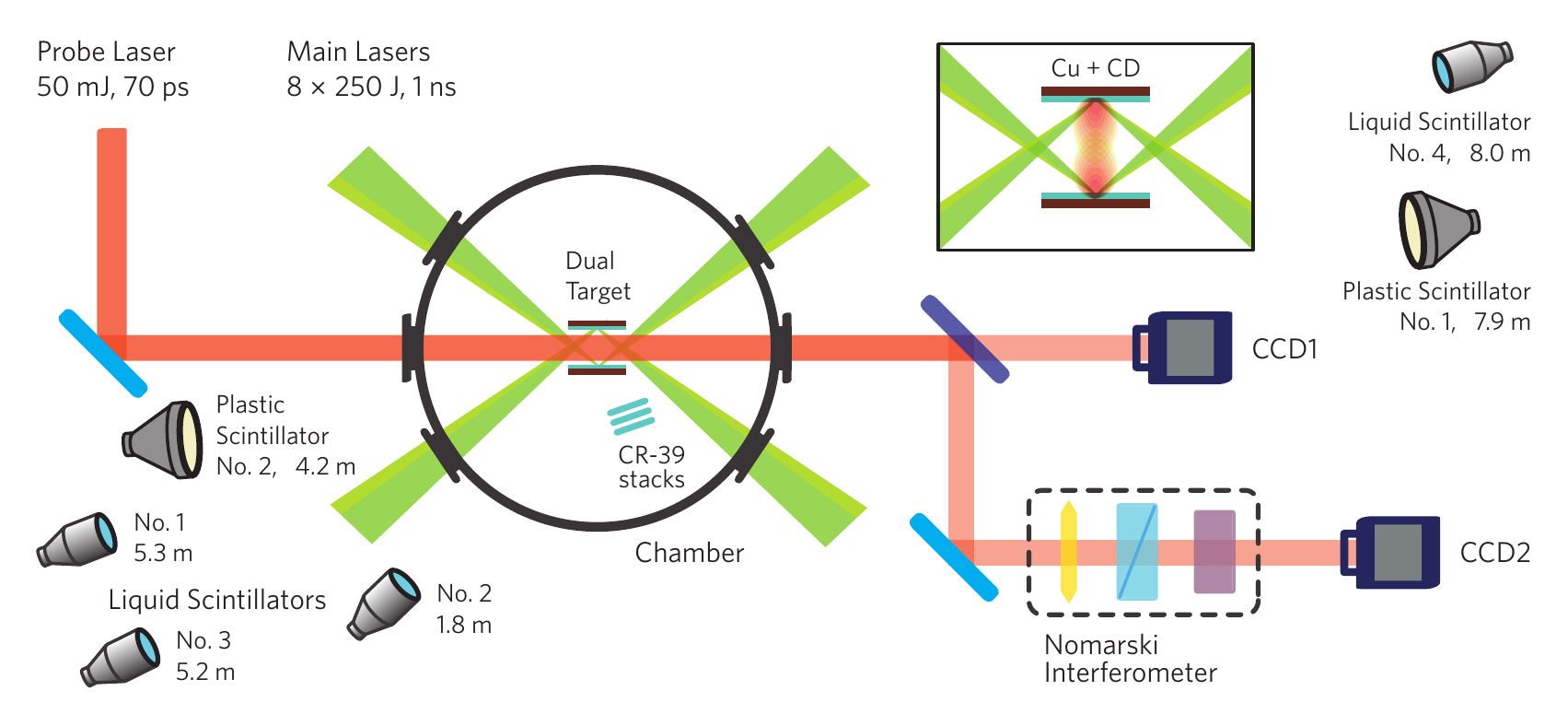}
 \caption{The experimental setup. Four laser beams were
 tuned to focused on the target at one side, and another 4 on the opposite side.
 By using the probe laser and a Nomarski interferometer, the optical images of the plasma were taken. 
 Neutron signals were recorded by scintillation detectors at different distances. }
 \label{fig.setup}
\end{figure*}

Another laser with a duration of 70 ps and wavelength of 526 nm was used as the probe, 
which passed through the plasma generated by the main laser beams. 
The interference images were taken by a Nomarski interferometer \cite{Abel-inv-Benattar1979, LiuXun2011-SG}. 
By tuning the delay time between the probe laser and the main lasers, snapshots of
the plasma at different times could be taken.

Neutron detectors were located outside the laser target chamber. 
Four of the them were liquid scintillation detectors 
(EJ-301) with scintillator size of $(\pi/4)\times 12.7^2\times $12.7 cm$^3$,
along with two plastic detectors (BC400) with scintillator size of
$(\pi/4)\times 25.4^2\times 5$ cm$^3$. All scintillators were directly
coupled with photomultiplier tubes (PMTs). The signals were recorded
by oscilloscopes with a bandwidth of 1 GHz. 
The neutrons were measured by the time-of-flight (TOF) approach.


Typical TOF spectra are shown in Fig.~\ref{fig.tof}, in
which the results from four liquid scintillation detectors at different 
locations are given. In each of the curves, the
first dip to the left, which saturates the detectors, represents the
photons induced by the high-intensity lasers. The photons, including
X-rays and $\gamma$-rays, are induced by the original 351 nm, 1 ns
width laser pulse and scattered secondary photons on the
materials around the targets. Since most X-ray and $\gamma$-ray
emissions in atoms or nuclei are in a smaller-than 1 ns domain, they
are expected to arrive at the detectors as a ns-width pulse, which
is the same pulse width as the original driving laser. Within such a
narrow width, the photons are highly overlapped, and some detectors
may be saturated. The long tail of the first dip is
due to the long discharging time of the PMTs.

The second dip in the curves in Fig.~\ref{fig.tof} represents
the neutron products. The neutrons from the D(d,n)$^3$He
reaction have an energy of $E_n=2.45$ MeV, or an speed of 2.16~cm/ns, 
which is much smaller than that of the photons (30~cm/ns).
The expected  neutron speed and the measured neutron speeds at different
detector locations show good agreement with each other.

\begin{figure}
 \centering
 \includegraphics[width=8cm]{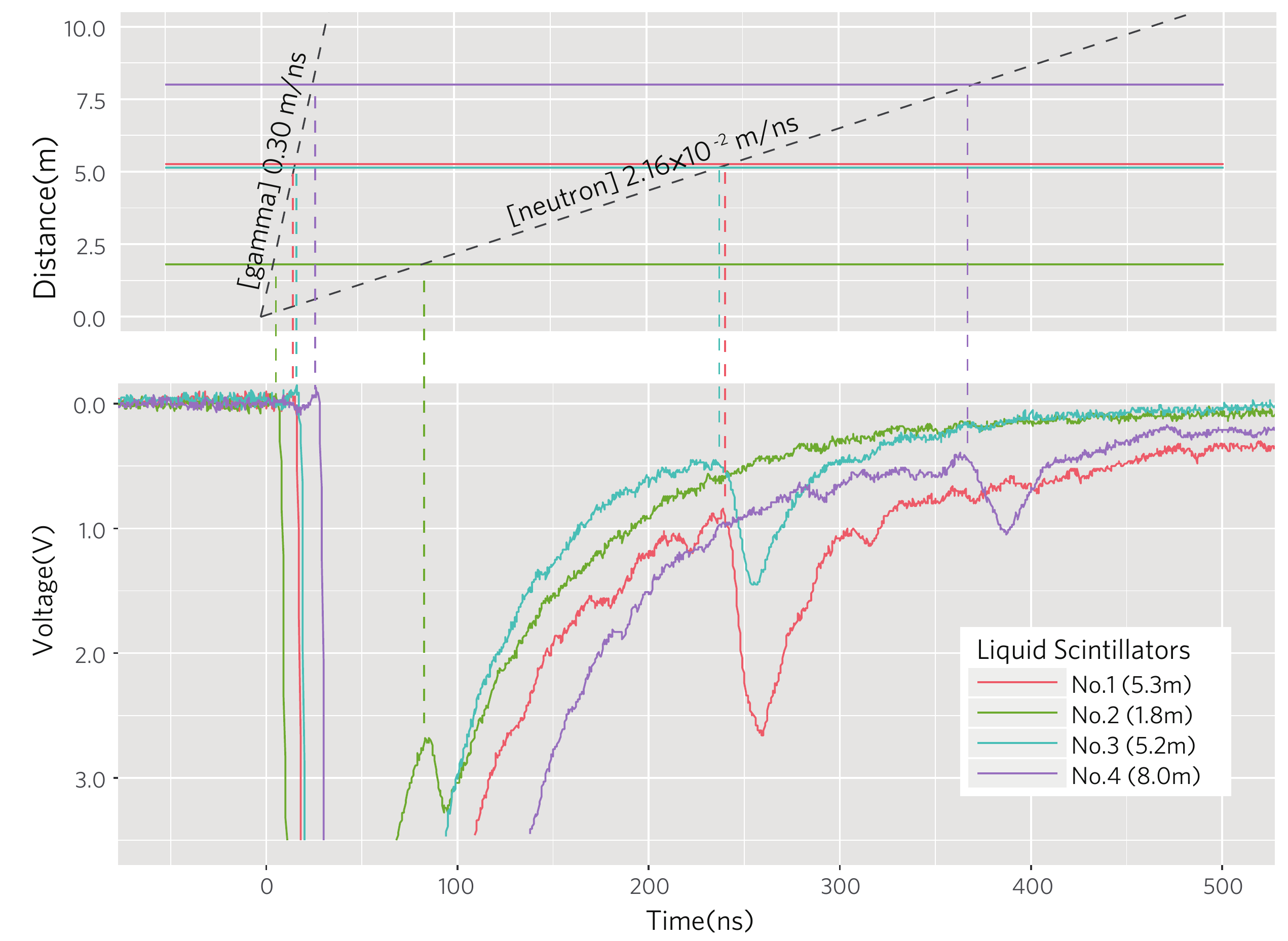}
 \caption{
 Typical time-of-flight (TOF) spectra of neutron detectors.
 Data from the four liquid scintillation detectors at distance 1.8 m, 5.2 m, 5.3m, and 8.0 m, are shown.
The expected arriving time of photons and neutrons (2.45 MeV) for each detector are indicated by dashed lines. 
They match well with the TOFs measured by detectors at different distances.
}
 \label{fig.tof}
\end{figure}

To obtain absolute neutron yields, all detectors have been calibrated by
using two radiation sources, a D-D neutron generator and a
$^{137}$Cs $\gamma$-ray source. The Monte Carlo (MC) simulation
code, GEANT4 \cite{Geant42003}, is employed for the calibration. 
In the MC simulation, after the to-be-detected particles depositing
energies in the scintillator, the energy is transferred
to luminescent photons with different efficiencies for neutrons and
$\gamma$-ray \cite{nCal-NIMA2014}. 

The final neutron yields for different runs
are obtained by combining the MC simulation, 
the experimental calibration data of the $^{137}$Cs $\gamma$ source and the neutron generator, 
as well as weighted values of all detectors.
The results are shown in Tab. \ref{results.tab}.

In the experiment, the observed neutrons might come from three
sources: from the original laser-induced fireballs ($N_{fb}$), from
the cold target when the energetic deuterium ions from the opposite target
hit it ($N_{cold}$), and from the area where two plasma currents
collide with each other ($N_{collide}$). The total neutron yield
is the sum of all the three sources, i.e.
$N_{total}=N_{fb}+N_{cold}+N_{collide}$.

To see how many neutrons come from the original fireballs, 
we either took off the second CD target (Tab. \ref{results.tab}, Run35), 
or left only a target base there but without CD film on it (Tab.
\ref{results.tab}, Run38--40), 
while keeping all other laser parameters the same. 
The results have shown no evidence of neutrons from those runs,
i.e. under the detecting limit of $<2\times10^{3}$ (95\% C.L.).

To determine $N_{cold}$, four lasers were focused on one of the two CD
targets, but no laser directly focused on the opposite one. 
The neutron yields for this setup was $0.5\times 10^{5}$ (Tab. \ref{results.tab}, Run80 and 81),
which was much smaller than the cases with double targets. 
It should be pointed out that the second target was not totally
``cold''. Since this target was only 4.4 mm away from the
opposite one, it could also be ionized by the scattered laser and the X-ray
coming from the opposite target \cite{NIF-X-ray-compress2012}. In fact, plasma on the 
opposite target surface was observed on the interferometer images in these runs.

Comparing the measured $N_{fb}$, $N_{cold}$, and $N_{collider}$, one
can conclude that the neutron yields are dominated by ion collisions, 
and the present setup provides an efficient way to
ignite nuclear fusion reactions in a mini-size collider.

The Abel inversion approach has been employed to deduce the electron
density $n_e$. Since the spacial distribution of the plasma is not ideally symmetrical,
we followed a numerical method for asymmetrical Abel inversion
described in Ref. \cite{yas81}. 

For the CD$_{1.29}$ targets used in the experiment,
considering the charge neutrality of plasma in the $\mu$m scale and
the fact that the carbon and deuteron atoms were fully ionized in the
energy range of interest ($>5$~keV), the density of deuterons in the
plasma can be estimated as
\begin{equation}
n_D\approx\frac{1.29}{6+1.29}n_e.
\end{equation}
The deuteron density estimated in this way has been confirmed 
by the MULTI2D hydrodynamics code \cite{Ramis2009}.

A typical density distribution is shown in Fig. \ref{fig.diff}. From
the data, one can estimate the collision frequency between electrons
(e-e), electrons and deuteron ions (e-D), and D-D.
The D-D mean free path, $\lambda_{DD}$, can be written as
\cite{chenais1997kinetic}:
\begin{equation}
\lambda_{DD}=\frac{m_D^2 v_{12}^4}{4\pi  Z^4 e^4\, n_D \ln \Lambda_{12}} ,
\end{equation}
where $m_D$ is the deuteron mass, $v_{12}$ the relative velocity,
$Ze$ the ion's change, and $\ln\Lambda_{12}$ the so-called ``Coulomb
logarithm'' \cite{ramazanov2001coulomb}. With our experimental
setup, in the relative velocity range $v_{12}>1\times 10^{8}$ cm/s,
(corresponding to $E_{cm}>5.2$ keV), $\lambda_{DD}$ is calculated to
be larger than 46 mm, which is much larger than the separation
between the two targets (4.4 mm). 
Therefore the plasma is really ``collisionless'' for deuteron energy larger than 5.2 keV. 
The e-D collision frequency of
$3\times10^{10}$/s can also be estimated. This means that the
electrons could collide with other electrons and ions for about 30
times in 1 ns. Consequently, a quasi thermal equilibrium of ions is
established.

\begin{figure}
 \centering
 \includegraphics[width=9cm]{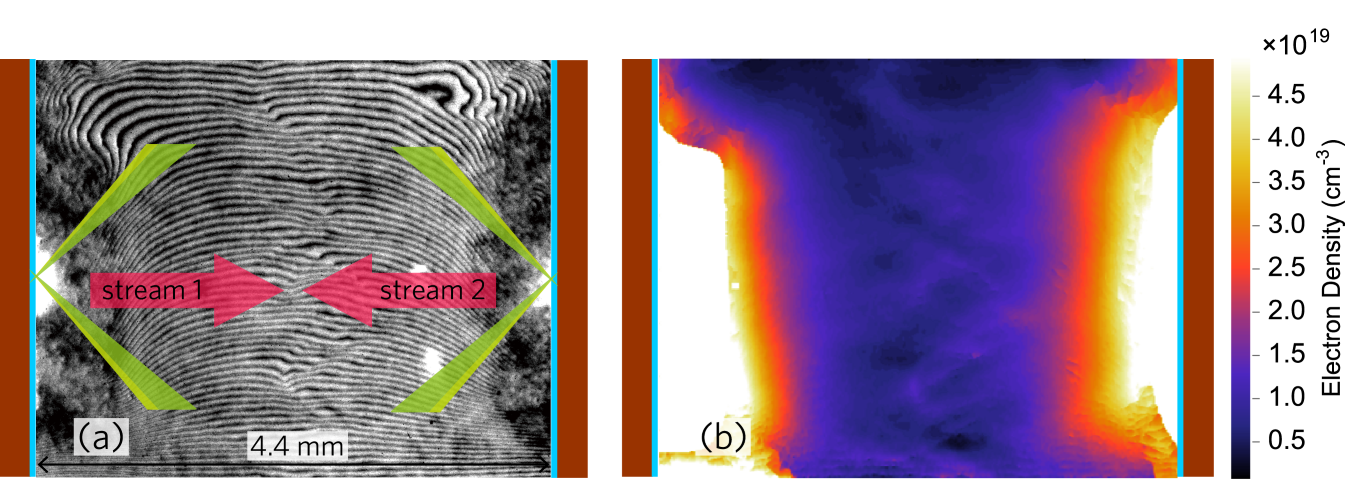}
 \caption{
 (a) A typical Nomarski interferogram of the plasma streams, 
and (b) the corresponding electron density distribution derived by Abel inversion approach.
}
 \label{fig.diff}
\end{figure}

A numerical calculation has been carried out with a simplified plasma dynamic model to obtain the  expected neutron yields. 
Considering the fact that $N_{fb}$ is very small, we assume that the deuterons from 
one target can only have reactions with those from the opposite side, 
and the D-D neutrons from the same side are negligible. 
The reaction yield can be written as \cite{rolfs1988cauldrons},
\begin{equation}\label{eq.yield}
Y=\iiint n_{1D}(\vec{r}_1)\ n_{2D}(\vec{r}_2)\sigma(v_1,v_2) \mathrm{d}\vec{r}_1\mathrm{d}\vec{r}_2 \mathrm{d}S,
\end{equation}
where $n_{1D}$ and $n_{2D}$ are the deuteron densities of the left
and right sides, respectively. 
The cross section is
\begin{equation}
\sigma(E_{\mathrm{cm}})=S(E_\mathrm{cm})\exp(-2\pi\eta)/E_\mathrm{cm},
\end{equation}
where $E_\mathrm{cm}=\frac{m}{4}(v_1+v_2)^2$ is the center-of-mass
energy (c.m.e.), $S(E_\mathrm{cm})$ defined by this equation is the so-called
astrophysical S-factor, and $\eta=\frac{Z_1Z_2e^2}{\hbar (v_1+v_2)}$
is the Sommerfeld parameter.

We have simplified the ion speed $v$ as a constant in the
collisionless regime, i.e. $v=z/t_0$, where $t_0$ is the delayed time of the probe laser, 
and $z$ is the distance to the target. $n_D$ is separated into left part and right
part which are originally from the left and right targets, i.e.
$n_D=n_{1D}+n_{2D}$.


\begin{figure}
 \centering
 \includegraphics[width=8cm]{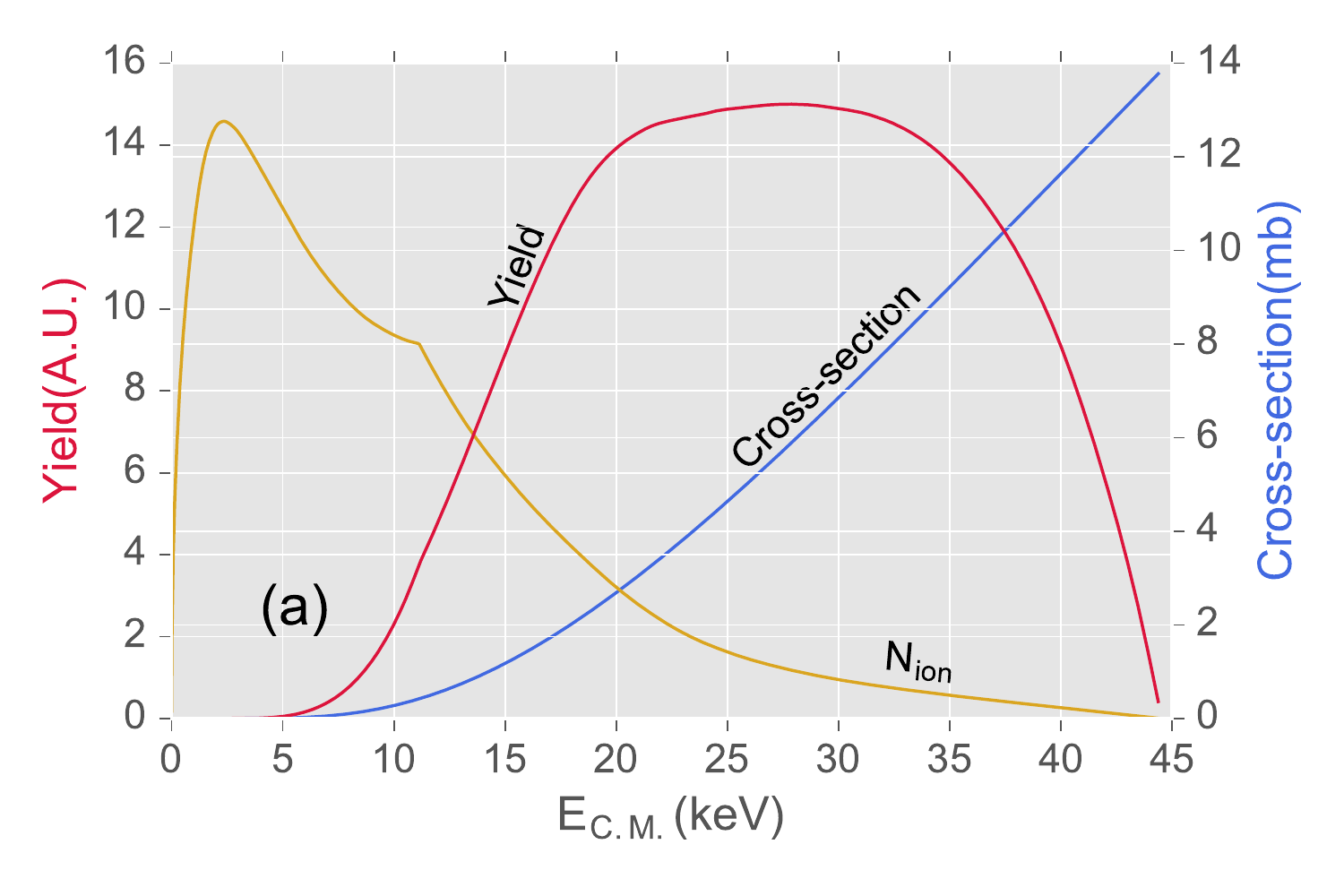}
 \includegraphics[width=7cm]{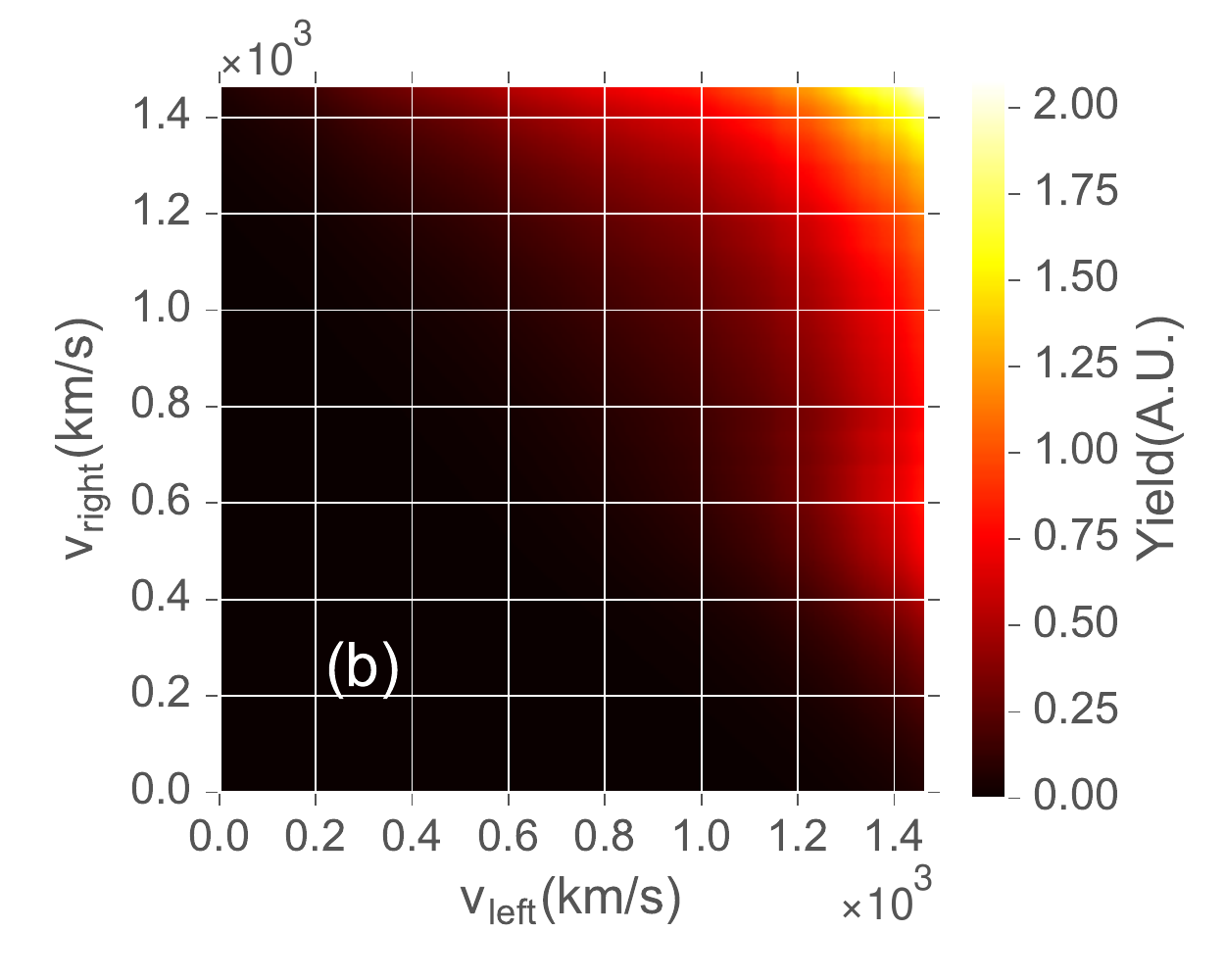}
 \caption{ 
Neutron yields contributed by deuterium ions with different c.m. energies. 
(a) the Gamow-window-like feature of the reaction yield.
Three lines are shown in this chart: 
the number of the ions, the D(d,n)$^3$He  cross section, and the neutron yield. 
(b) The neutron yield generated by ion pairs with different speed. The x-axis is the speed of the ions from the left target (V$_{left}$), and the y-axis, the right target (V$_{right}$).
}
 \label{fig.yield}
\end{figure}

Figure \ref{fig.yield} shows the D-D reaction yield for different
velocities in plasma with the assumptions described above. 
One can find that the  $N_{ion}(E_{cm})$, the number of ion pairs with
c.m.e. $E_{cm}$, decreases as $E_{cm}$ increases; 
while the  $\sigma(E_{cm})$ increases as $E_{cm}$ increases. 
Therefore, according to Eq. \ref{eq.yield}, large
reaction yields are found in the c.m.e. range of 15 to 40
keV, as shown in Fig. \ref{fig.yield}(a). This Gamow-window-like
\cite{rolfs1988cauldrons} structure implies that the present method
could be a promising new tool for the key reactions with
nuclear-astrophysical interest, which are otherwise very difficult,
if not impossible, to preform in traditional experimental setups.

The calculated neutron yield by using the simplified model is
$(3.1\pm1.2)\times10^6$ for Run37, or 8 times larger than the experimental
observation.
This disagreement may be due to the neglected self-generated
magnetic fields in our calculation.  
With the current experimental conditions, the head-on-head
collision of plasma streams can generate a toroidal field
inversely proportional to electron density and the distance to the
symmetrical axis, i.e. $B_\phi/n_e r$ = const
\cite{ryutov2013magnetic}. This type of fields has been reported in
previous experiments with the similar head-on-head collision setup
\cite{JetCollisionPRL2013, huntington2015observation}, 
and the magnetic filed strength is estimated to be about 10 T level. 
Under this field strength, the deuterons with an energy of tens keV can
be significantly bent from a straight path. 
Therefore the c.m.e. of the ion pairs should become smaller, 
thus resulting a smaller neutron yield.

\begin{table*}
\begin{ruledtabular}
\caption {The neutron yields at different runs. (See the text for details)}
\label{results.tab}
\begin{tabular}{c c c c c c c}
Run\# &	Lasers\footnote{``$M+N$'' means: $M$ lasers focused on the first target, and $N$ lasers on the second one. See the text for details.} 
&Laser Energy(J)   &Target1&Target2& Neutron yield \\
\hline
34 &    4+4      & 297$\times 8$     & CD(10$\mu$m)         &CD(10$\mu$m)   &	$(3.8\pm1.1)\times10^{5}$ \\
35  &    4+0     & 260$\times 4$   & CD(10$\mu$m)        &-     & 0 \\
36 &    4+4      &254$\times 8$    & CD(10$\mu$m)         &CD(10$\mu$m)   &	$(4.0\pm1.2)\times10^{5}$\\
37 &    4+4      &244$\times 8$    & CD(10$\mu$m)        &CD(10$\mu$m)   &	$(3.9\pm1.1)\times10^{5}$\\
38  &    4+4     &254$\times 8$    & CD(10$\mu$m)        &Cu(only)   &	0 \\
39  &    4+4     &217$\times 8$     & CD(10$\mu$m)        &Cu(only)   &	0 \\
40  &    4+4     &246$\times 8$    & CD(10$\mu$m)        &Cu(only)   &	0 \\
79  &    4+4     &230$\times 8$     & CD(10$\mu$m)      &CD(10$\mu$m)   &	$(1.0\pm0.5)\times10^{5}$\\
80  &    4+0     &263$\times 4$     & CD(10$\mu$m)     &CD(10$\mu$m)   &	$(0.5\pm0.3)\times10^{5}$\\
81  &    4+0     &219$\times 4$    & CD(10$\mu$m)     &CD(10$\mu$m)   &	$(0.6\pm0.3)\times10^{5}$\\


\end{tabular} 
\end{ruledtabular}
\end{table*}


The method introduced in this article could have far-reaching
implications for future nuclear reaction experiments aiming at
understanding of the origin of element production in the universe.

First, this method provides a controllable way to trigger nuclear
reactions within the Gamow window \cite{rolfs1988cauldrons}. 
Nuclear astrophysical interested reactions are normally have very small cross sections, and traditional accelerators have very low peak beam intensities, which result in the very low signal to background noise ratio. 
Therefore, it is very difficulty to study nuclear reactions in their Gamow windows with traditional accelerators. 
Compared with the coulomb explosion method \cite{DDn-Nature1999}, whose energy is hardly tunable, 
the current setup is likely the only known controllable method which can provide particles with quasi-Boltzmann distribution for nuclear astrophysical studies up to date.

Second, this setup provides a lab-based full plasma environment for nuclear reactions studies. 
There have been strong indications that the decay
properties and reaction rates of bare nuclei in plasmas differ
significantly from those of atomically-bound states obtained from
normal conditions\cite{Beta-decay2011,DD-e-Screening2001}. 
Compared with other methods, for example the
Coulomb explosion method \cite{DDn-Nature1999} or the storage ring
method \cite{Bertulani1997}, the plasma in this work is charge neutral, 
and the environment created here is more similar as that in real astrophysical
cases.

Third, the results show hints that
the self-generated macro-scale electromagnetic field may play an
important role in the nucleosynthesis of our universe.
A plasma in a ``collision'' state means that ions in it collide
with each other frequently, so that the system can quickly reach a
thermal equilibrium; while a ``collisionless'' state means that the
ions rarely collide and the system is far from a thermal
equilibrium. Because of the collisionless features, the nuclei
inside CLPs can process in macro-scale lengths before
being scattered, and thus acquire energy due to the self-generated
macro-scale electrical field \cite{B-Self-org-kugland2012}. 
This can accelerate the nuclei and/or change their paths, and therefore the
reaction yields could be completely different from that in
thermal equilibriums, as observed in the present experiment. 
More nuclear reaction studies in a laboratory CLP may an important step toward
solving the long-standing puzzles like the $^{6,7}$Li abundance in
Big Bang Nucleosynthesis \cite{cyburt2016BBN}.


In summary, we have presented a novel experimental method for
studying nuclear reactions in plasma environments. 
Taking advantages of the
extremely-high peak of ion flux and high temperature induced by
lasers, stellar environments could be simulated, in which low
energy and small cross section  nuclear reactions could be studied on
earth-based laboratories. By employing this method, the D(d,n)$^3$He 
reaction has been performed by using head-on-head collision of
plasma streams driven by nanosecond pulse lasers. 
The experimental results have shown that the neutron
yield from the nuclear reaction are enhanced significantly by the
head-on-head collision in the CLPs. 
And we have found evidences that the
self-generated electromagnetic fields in CLPs might affect the
nuclear reaction yield significantly. 
This novel plasma-version
mini-collider can be employed in future studies of nuclear
astrophysical processes, the magnetic confinement fusion, and possibly more others.


We would like to acknowledge the SG-II staff for operating the laser
facility, CAEP staff for the target fabrication. 
This work is supported by 
the National Basic Research Program of China (Grant Nos. 2013CBA01501, 2013CB834401), 
the National Nature Science Foundation of China (Grant Nos. 11135012, 11135005, 11375114), 
and the Global R\&D Networking Program funded by the Republic of Korea's Ministry of Science, ICT and Future Planning (Grant No. NRF-2012-0004839).
One of us (CBF) thanks for the supports 
from Shanghai Municipal Science and Technology Commission (under grant No. 11DZ2260700)), 
the Key Lab for Particle Physics, Astrophysics and Cosmology, Ministry of Education, 
and Shanghai Key Lab for Particle Physics and Cosmology(SKLPPC).



%

\end{document}